\documentclass[twocolumn]{aa}
\usepackage{graphicx}
\usepackage{txfonts}
\usepackage{natbib}

\begin{document}

\title{Numerical simulation of fundamental trapped  sausage modes}

\author{M. C\'ecere\inst{1} \fnmsep\thanks{CM: Mariana C\'ecere, cecere@famaf.unc.edu.ar} 
\and 
A. Costa\inst{1,2} 
\and 
O. Reula\inst{1,4}}

\institute{Facultad de Matem\'atica, Astronom\'\i a y F\'\i sica (FaMAF, UNC), C\'ordoba, 5000, Argentina     \and
Instituto de Astronom\'{\i}a Te\'orica y Experimental (IATE, UNC-CONICET), C\'ordoba, 5000, Argentina     \and
Facultad de Ciencias Exactas, F\'\i sica y Naturales (FEFyN, UNC), C\'ordoba, 5000, Argentina     \and Instituto de F\'\i sica Enrique Gaviola (IFEG, CONICET-UNC), C\'ordoba, 5000, Argentina}
  
\date{Received ; accepted }

\abstract
{We integrate the 2D MHD ideal equations of a straight slab to simulate observational results associated with fundamental sausage trapped modes.} {Starting from a non--equilibrium state with a dense chromospheric layer, we analyse the  evolution of the internal plasma dynamics of  magnetic loops, subject to line--tying boundary conditions, and with the coronal parameters  described in Asai et al. (2001) and Melnikov et al. (2002) to investigate the onset and damping of sausage modes.} {To integrate the  equations we used   a high resolution shock--capturing (HRSC) method  specially designed to deal appropriately with  flow discontinuities. }{Due to non--linearities and inhomogeneities, pure modes are difficult to sustain and always occur  coupled among them so as to satisfy, e.g., the line--tying constraint. 
We found that, in one case, the resonant coupling of the sausage fundamental  mode with  a slow one  results in a non--dissipative  damping of the former. }
{In scenarios of thick and dense loops,  where  the analytical theory predicts the existence of fundamental trapped sausage modes, the coupling of fast and slow quasi--periodic modes -with a node at the center of the longitudinal speed- occur contributing to the damping of the fast mode. If a discontinuity in the total pressure  between the  loop and the corona is assumed, a fundamental fast sausage transitory leaky regime is spontaneously produced and  
an external compressional Alfv\'en wave takes away the magnetic energy.}


\maketitle

\section{Introduction}

Coronal heating  remains  today an open field  of research. One of the most important contribution to it is attributed to active region overdense loops which are bright in soft X--ray and EUV (Aschwanden et al. \citealp{ac10}).
Over the past decade the interest in determining the presence of waves and oscillations that can contribute to the heating mechanisms has increased on the basis of the development of theoretical and observational studies.  The  coronal seismological remote diagnostics revealed an effective media to identify a wide--spectrum of fast and slow intensity oscillations. Measurements of characteristic periods, dispersion relations,  speeds and damping times as well as different theoretical models that intend to describe non-dissipative damping mechanisms and leakage processes of systems with high Reynolds number  provide  new remote sensing diagnostic tools to reveal unknown or more accurate solar physics parameters, e.g.,  magnetic field strength and transport coefficients (see, e.g. Aschwanden \citealp{asc4} and Nakariakov and Verwichte \citealp{nak}, for recent reviews).

The pure sausage mode, modelled as  a magnetic cylinder or a magnetic slab, is known to be a compressible MHD fast  mode  that perturb the plasma in the radial direction  causing a symmetric contraction and widening of the tube without distortion of its axis (Edwin and Roberts \citealp{rob}). This  mode  was  observationally detected  in  the corona by Nakariakov et al.  \cite{nak2}  using  the Nobeyama Radioheliograph. EUV and soft X--ray band periodic variations of the thermal emission intensity are associated with density perturbations and Doppler broadening of the emission lines. Zaitsev and Stepanov \cite{zai} proposed that a modulation of hard X--ray and white--light emission from the loop footpoints can occur due to the change in the loop radius and the consequent change 
of the mirror ratio, causing the periodic precipitation of non--thermal electrons of flaring loops. These quasi--periodic pulsations, triggered by sausage modes, were also associated  with flaring events of other stars (Mitra--Kraev et al. \citealp{mit}). 

The standard theoretical models (Edwin and Roberts \citealp{rob};  \citealp{rob2};  Zaitsev and Stepanov \citealp{zai}) predict that 
the fast global sausage mode  has two regimes separated by a cutoff at the long wavenumbers: the leaky one, where the energy is lost through the boundaries of the loop and carried away by an external solution, occurring in thin and long loops when the internal Alfv\'en speed is larger than the external one; and  the trapped regime, occurring in thick and short loops, where the energy of the oscillation is confined to the interior of the loop structure.
Aschwanden et al. \cite{asc} showed  that the sausage cutoff imposes such a high electron density that they can only occur in flare loops. They found that previously reported fast global mode observations had oscillations confined to a small segment of the loop with nodes limiting the segment, indicating that the oscillation must be associated with higher harmonic modes  rather than fundamental ones. However, they found that  microwave and soft X--ray observations by Asai et al. \cite{asa} and Melnikov et al. \cite{mel} are consistent with fast sausage MHD oscillations at the fundamental harmonic.
 The leaky regime has recently been studied by Pascoe et al. \cite{pas}, who found by  numerical modeling that long loops with sufficiently small density contrast can support global sausage leaky  modes of detectable quality, and thus they argue that this mode can also be responsible of the quasi--periodic pulsations of flaring emission. An important issue is that the global sausage leaky mode period is determined by the Alfv\'en speed outside the loop. Also, Verwichte et al. \cite{ver} noted that wave energy of loops  with internal Alfv\'en speed locally below the external one can tunnel through an evanescent barrier into the surrounding corona 
due to the inhomogeneity of the media that produces relative variations of the Alfv\'en speeds. Recently, Srivastava et al. \cite{sri} based on the observation of multiple sausage oscillations  suggested that  fundamental post flare cool loops are not in equilibrium causing plasma motion along the loop.  Pascoe et al. \cite{pas2} investigate the effects of non--uniform cross--sections and found that the global sausage mode is very effectively excited in flaring loops. 

The sound $(v_{s})$ and Alfv\'en $(v_{A})$  speeds are associated with  two main properties of the low corona, its compressibility and  elasticity, respectively.  The low coronal values of the parameter $\beta \simeq v_{s}^{2}/v_{A}^{2}<<1,$  imply  that the sound speed is much smaller than the Alfv\'en one.  Due to the high Reynolds number in the corona and to the rapid  fast mode  decays, this must be firstly accomplished in an almost non--dissipative way. When more realistic inhomogeneous distributions of the physical quantities are taken into account, both in the longitudinal and radial direction (e.g., dense chromospheric layer), the pure character of the MHD modes is lost and the nonlinear coupling of modes  takes place. 
In mediums with usual strong  sudden energy release sources, the drain of the larger Alfv\'enic elastic energy to the lower magnetoacoustic compressive perturbations 
would not be unusual, leading to the eventual non--linear steepening  of modes  and to   the formation of coronal shocks (Costa et al. \citealp{cos}). The quantitative evaluation of the statistical distribution and energy content in shock waves is important to estimate its contribution as a source for the coronal heating. 

In this paper our  aim is to investigate the onset and damping of coronal sausage modes considering nonlinear and inhomogeneous initial states. We simulate two studied observational flaring loops supposed capable to develop global sausage trapped modes, described in  Asai et al.  \cite{asa} (Case A) and Melnikov et al.  \cite{mel} (Case M), respectively. 
\begin{figure}
\begin{center}
 \includegraphics[width=7.4cm]{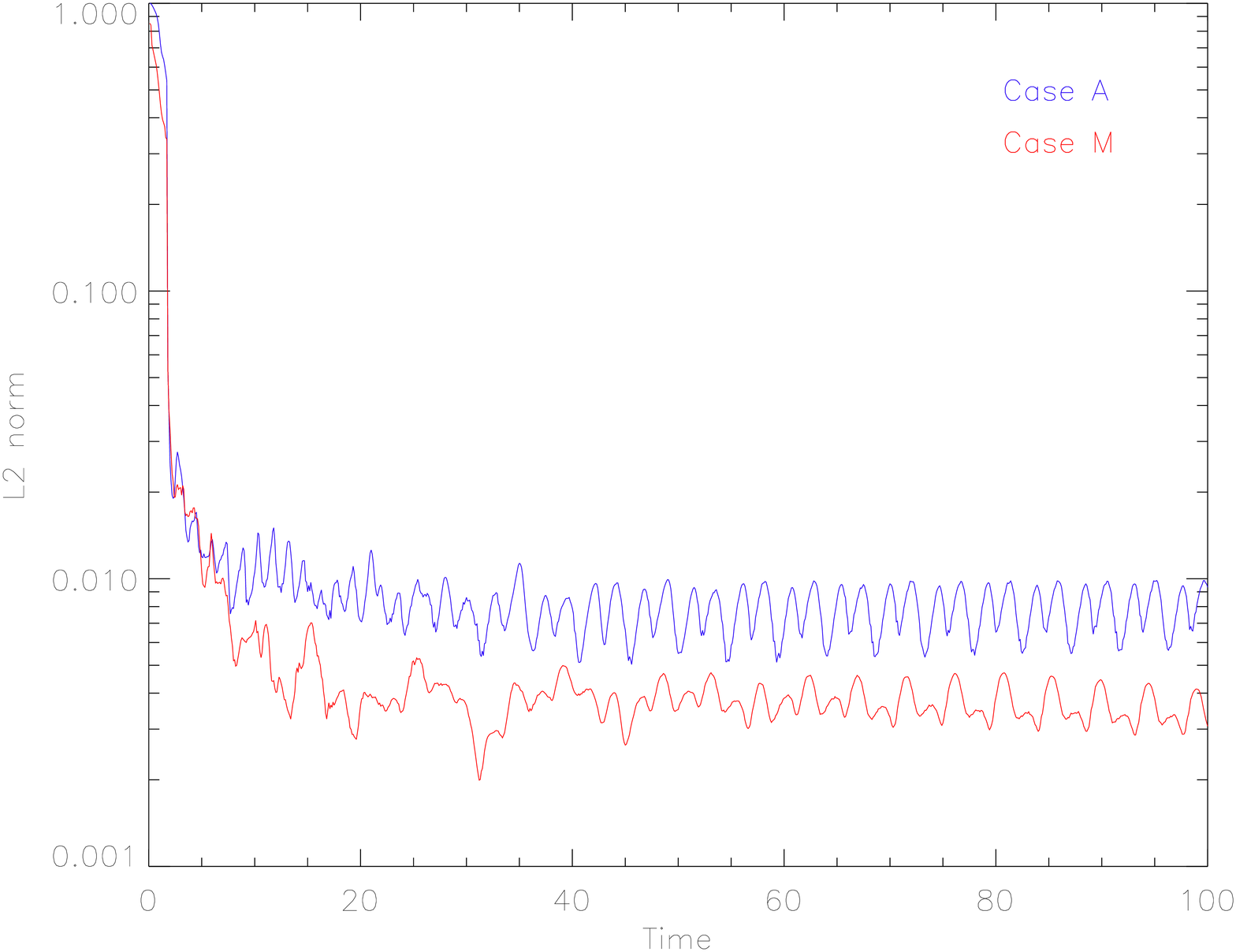}
 \end{center}
\caption{Divergence cleaning test. Dimensional time obtained multiplying by $t_{0}=1.39$sec for Case A and $t_{0}=3.25$sec for Case M.}
  \label{fig:cero}
\end{figure}

\begin{figure*}
\begin{center}
 \includegraphics[width=7.4cm]{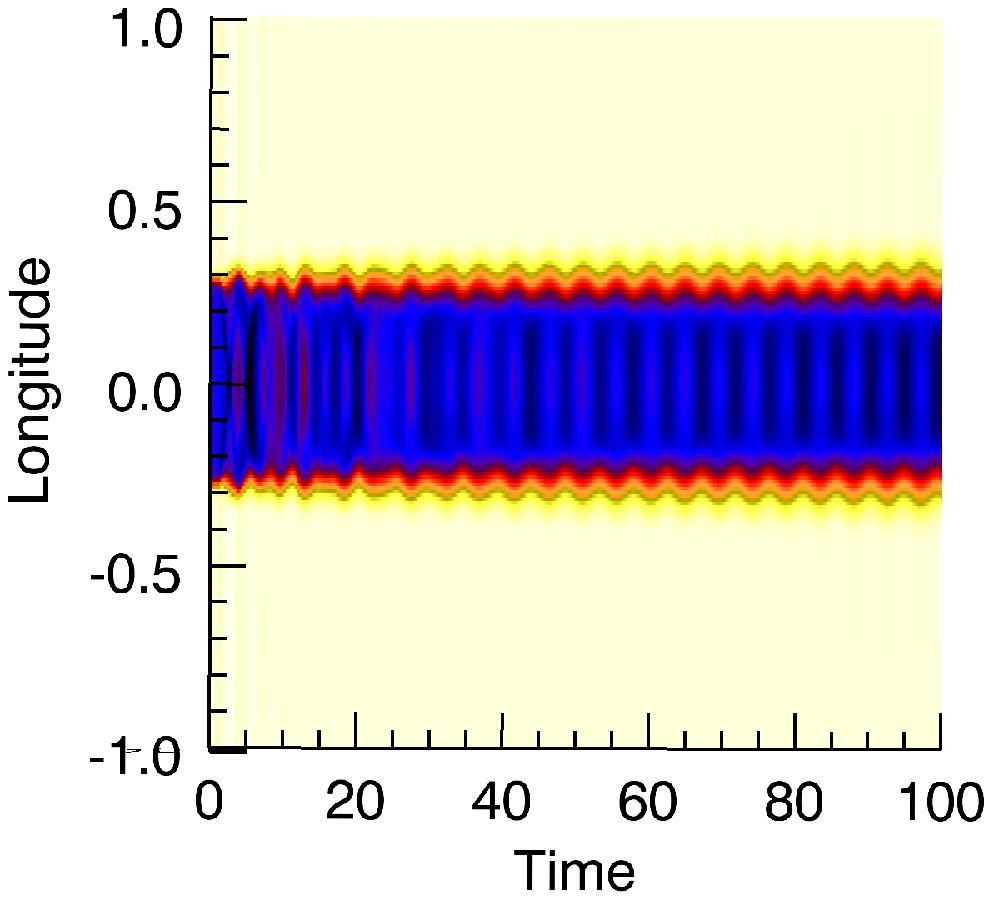}
 \hspace*{19pt}
  \includegraphics[width=7.4cm]{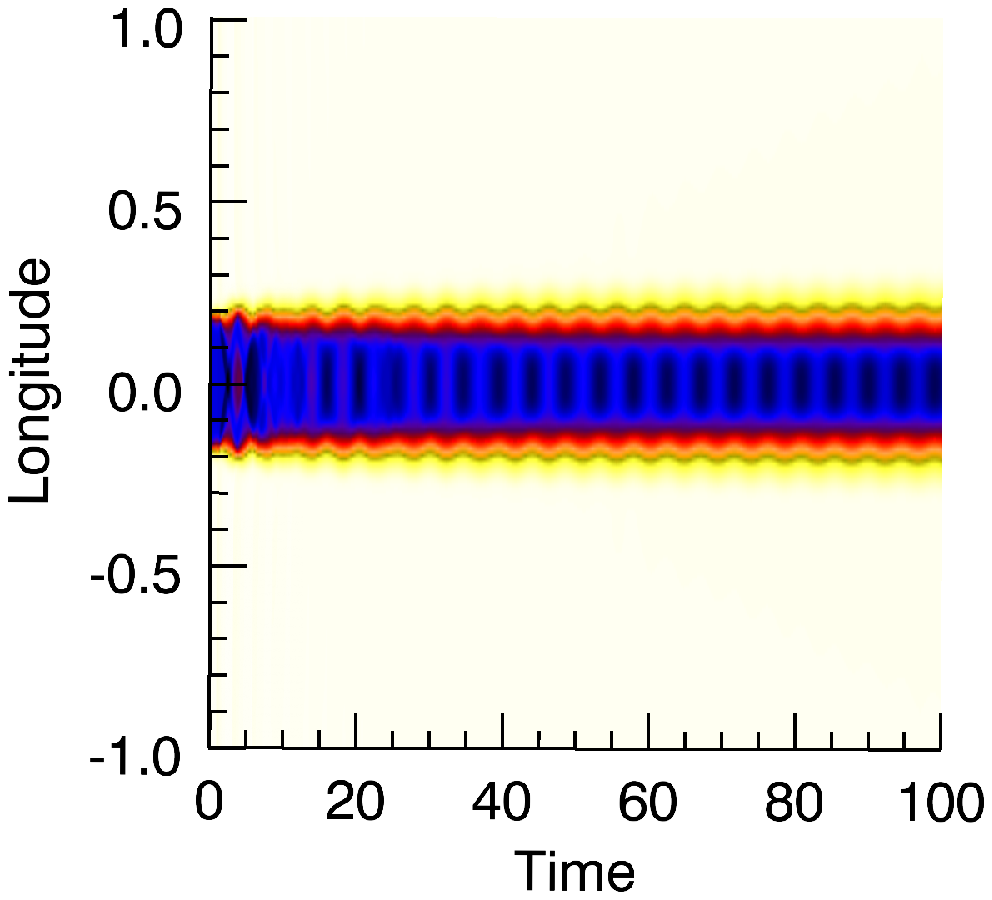}
 \end{center}
\caption{Density as a function of time, across the loop width ($x$), and at the center of the loop length, ($y=0$). Left) Case A, dimensional values obtained multiplying by $l_{0}=10$Mm and $t_{0}=1.39$sec; Right) Case M, dimensional values obtained multiplying by $l_{0}=16.3$Mm and $t_{0}=3.25$sec. }
  \label{fig:uno}
\end{figure*}

\begin{table*}
\begin{center}
\begin{tabular}{|l|c|c|c|c|c|c|} \hline
\multicolumn{1}{|c|}{\bf Regions} & 
\multicolumn{1}{c|}{$n_{A}$[$10^{10}$ \ $cm^{-3}$]\rule{0pt}{9pt}} &
\multicolumn{1}{c|}{$T_{A}$[$MK$]}& 
\multicolumn{1}{c|}{$B_{A}$[$G$]}& 
\multicolumn{1}{c|}{$n_{M}$[$10^{10} \ cm^{-3}$]} &
\multicolumn{1}{c|}{$T_{M}$[$MK$]} &
\multicolumn{1}{c|}{$B_{M}$[$G$]} 
\\ \hline
Inner ($0$\rule{0pt}{9pt})        & $4,3$    & $2.4  $ & $152$  & $9.8$   & $9$   & $90$    \\ \hline
External ($e$\rule{0pt}{9pt})     & $0.21$   & $2.5  $ & $153$  & $0.16$  & $8.8$ & $104$   \\ \hline
Chromosphere ($c$\rule{0pt}{9pt}) & $19000$  & $0.02 $ & $106$  & $1900$  & $0.1$ & $72$    \\ \hline
\end{tabular}
\end{center}
\caption{\label{tab:table1} Initial parameter values: particle number density $n,$   temperature $T,$  magnetic field $B$.  Left) Case A values taken from Asai et al. \cite{asa}; Right) Case M values taken from Melnikov et al. \cite{mel}.   }
\end{table*}
\section{Numerical procedure}

The non-linear nature of MHD equations, implies that solutions for complex systems must be obtained by numerical means and a suitable numerical implementation must be constructed for this purpose. Even  smooth initial data can easily lead to discontinuities or shock wave solutions which are not well resolved by standard  finite difference methods and  break down when the solution is not smooth enough.
To simulate the plasma  evolution of low coronal loops we used a numerical code specially designed to capture discontinuities via  the implementation of a high resolution shock-capturing (HRSC), methods together with  a Runge-Kutta method to evolve in time (Palenzuela et al.  \citealp{pal}).

The  initial boundary conditions  represent a coronal loop as a straight slab of enhanced density with footpoints in a dense chromospheric region and a magnetic field parallel to the axis of the slab. Gravity effects are neglected for simplicity and because, for the moment, we are not interested on curvature effects. Also the pressure scale height is $\sim 10^{2}$Mm. 
The non--dimensional  ideal MHD equations simulated by the codes are
\noindent 
$$\partial_{t} \rho =\partial_j (\rho v_j)$$
$$\partial_{t} (\rho v_i) = -\partial_j \left[ \rho v_j v_i + \left(p+\frac{B^{2}}{2}\right) \delta_{ij} - B_j B_i \right] $$
$$\partial_{t} e = -\partial_j \left[ \left( e+p+\frac{1}{2}B^2\right) v_j -  v_i B_i B_j\right] $$
$$\partial_{t} B_i =-\partial_{j}  (v_j B_i - B_j v_i) - \partial_{i} \phi $$
$$\partial_{t} \phi= -\partial_i B_i  $$
where
$$e = \frac{p}{\gamma -1} + \frac{1}{2}\rho v^2 + \frac{1}{2} B^2$$
the energy density, $\rho$ is the mass density, $ \mathbf{v}$ is the velocity, $\mathbf{B}$ is the magnetic field, $p$ is the thermal pressure and $\gamma=5/3$ is the specific heat ratio. In addition there is a fictitious variable $\phi$, coupled to the system, whose time evolution keeps under control the behavior of the magnetic field divergence. The line--tying of the magnetic field lines at the chromospheric basis ($y=\pm L/2,$ $L$ the length of the loop) is assured by choosing appropriate parameters that resemble the physical conditions, i.e. the change in the plasma parameter $\beta.$  

The condition for the existence of a global sausage mode  (Aschwanden et al. \citealp{asc}) is
\begin{equation}
\frac{L}{\omega}<0.65 \sqrt{\frac{\rho_{0}}{\rho_{e}}},  \ \ \ \ \ \ \
\label{1}
\end{equation}
$\omega$ the width of the loop, $\rho_0$ and $\rho_e$ the internal and external to the loop densities, respectively.
The global stationary period must satisfy the relation:
\begin{equation}
\frac{2L}{v_{Ae}}<P_{saus}=\frac{2L}{v_{ph}}< \frac{2L}{v_{A0}},  \ \ \ \ \ \ \
\label{2}
\end{equation}
where $v_{A0}$ the internal Alfv\'en speed, $ v_{ph}$ the phase speed and, $v_{Ae}$ the external Alfv\'en speed. 

For both observational cases we used a $200^2$ grid of  size $x \times y=$ $2$x$2,$ corresponding to $l_0 = 10$ Mm unities of simulation in Case A,  and to $l_0 = 16.3$ Mm in Case M. The reference values used to obtain  the non--dimensional equations are: 
$t_{0}=1.39$sec, $\rho_0 = 3.56 \ 10^{-15}$g cm$^{-3},$  $ B_0 = 152$ G, $v_0 = 7194$km s$^{-1};$ for Case A, and  $t_{0}=3.25$sec, $\rho_0 = 2.72 \ 10^{-15}$g cm$^{-3},$  $ B_0 = 90$ G, $v_0 = 5000$km s$^{-1},$ for Case M. The pressure reference value is calculated as $ p_0 = \rho_0  \ v_0^{2}. $
The loop length, the width and the inner density of the loops  were taken in accordance with the observational results. In the stationary stage, Case A loop has a  $L=15$ Mm length and a $\omega=5.6$ Mm  width; and Case M has a  $L=24.5$ Mm length and a $\omega=5.8$ Mm width. Table~\ref{tab:table1} shows the initial parameter values for the two trapped modes (eq.1). 

\subsection{Divergence cleaning test}

As noticed above we use a divergence cleaning scheme (Dedner et al. \citealp{ded}). We start with initial data which does not satisfy 
$\nabla\cdot  \mathbf{B}=0  .$ However the  departure of the calculation  from this constraint  can be evaluated through the ratio
\begin{equation}
\frac{\parallel \nabla\cdot  \mathbf{B} \parallel_{L2}}{\parallel  \mathbf{B} \parallel_{H1}}, 
\label{3}
\end{equation}
where
\noindent
$$
(\parallel  \mathbf{B} \parallel_{H1})^{2}=\int_{V}  \left( {B_{x,x}}^{2}+  {B_{x,y}}^{2}+{ B_{y,x}}^{2}+ {B_{y,y}}^{2}\right) dV$$
$B_{i,j}$ indicates the derivative of $B_{i}$ with respect to $j.$
We perform this test for the two cases.  Figure~\ref{fig:cero} shows that in five adimensional time units the  evolution drives the solution to a very small 
constraint violation, $\leq 1\%.$ 

\section{Results and discussion}
\subsection{Total equilibrium pressure across the loop radius}

We first considered total equilibrium pressure between some  interfaces, namely chromosphere -- inner loop region and inner loop region --  outer loop region. 
In the two  cases, the line--tied loop system evolves  spontaneously, in few periods ($\leq  4$)  to a stationary regime.  In the transitory, a density perturbation with energy draining into the exterior takes place  until the sausage trapped regime is established. 
Figures~\ref{fig:uno}a,b  show the density perturbation at the loop's  center and across its transverse size ($x$), as a function of time for Case A and M, respectively. Case M is appreciable damped in $\sim 9$ periods. The sausage mode   modifies  the radius   with a fundamental  amplitude of  $\sim 15\% R_{l},$  ($ R_{l}$  the loop radius) in Case A and $\sim 17\% R_{l},$  in Case M.

\begin{figure*}
\begin{center}
 \includegraphics[width=7.5cm]{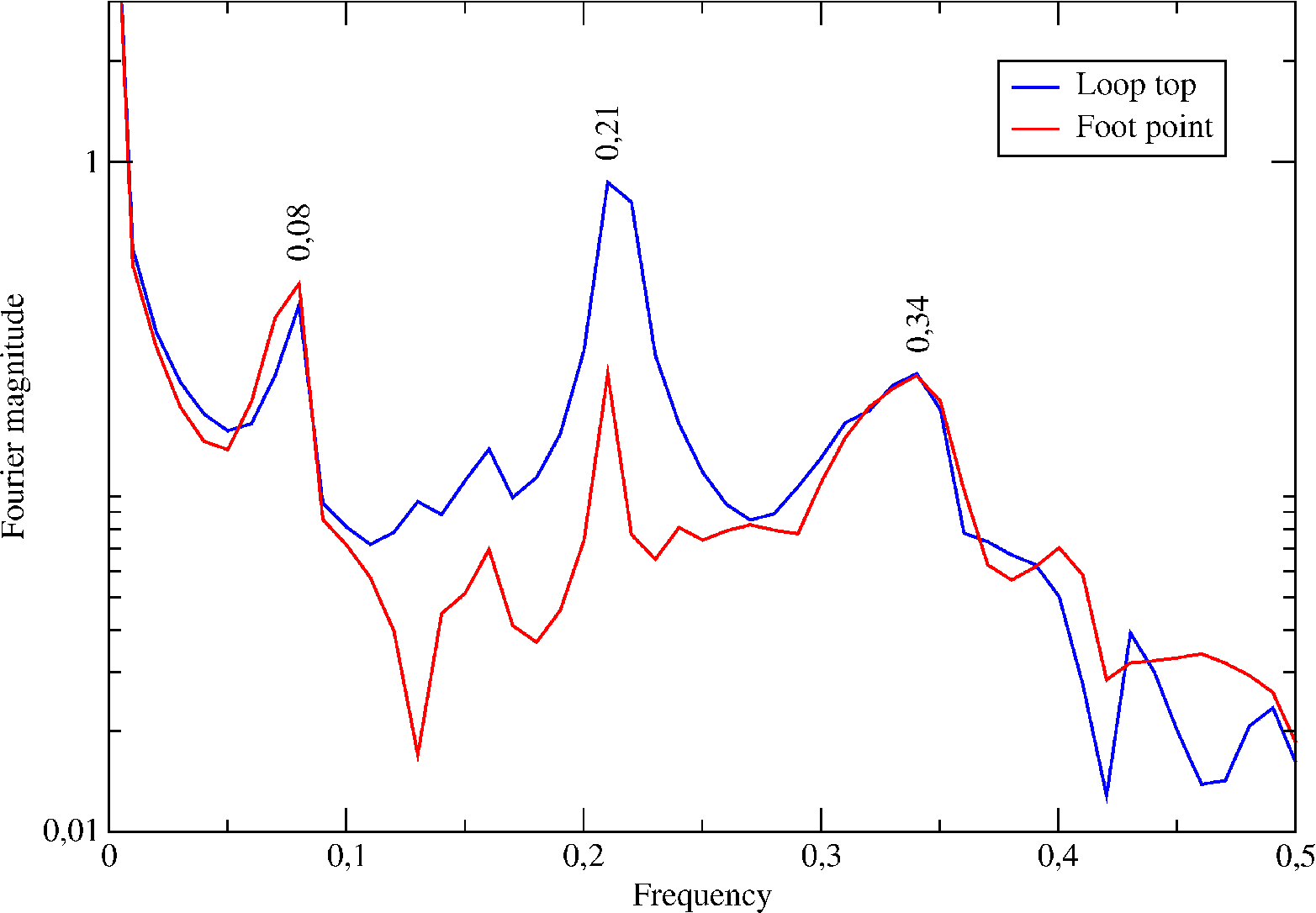}
   \hspace*{60pt}
 \includegraphics[width=7.5cm]{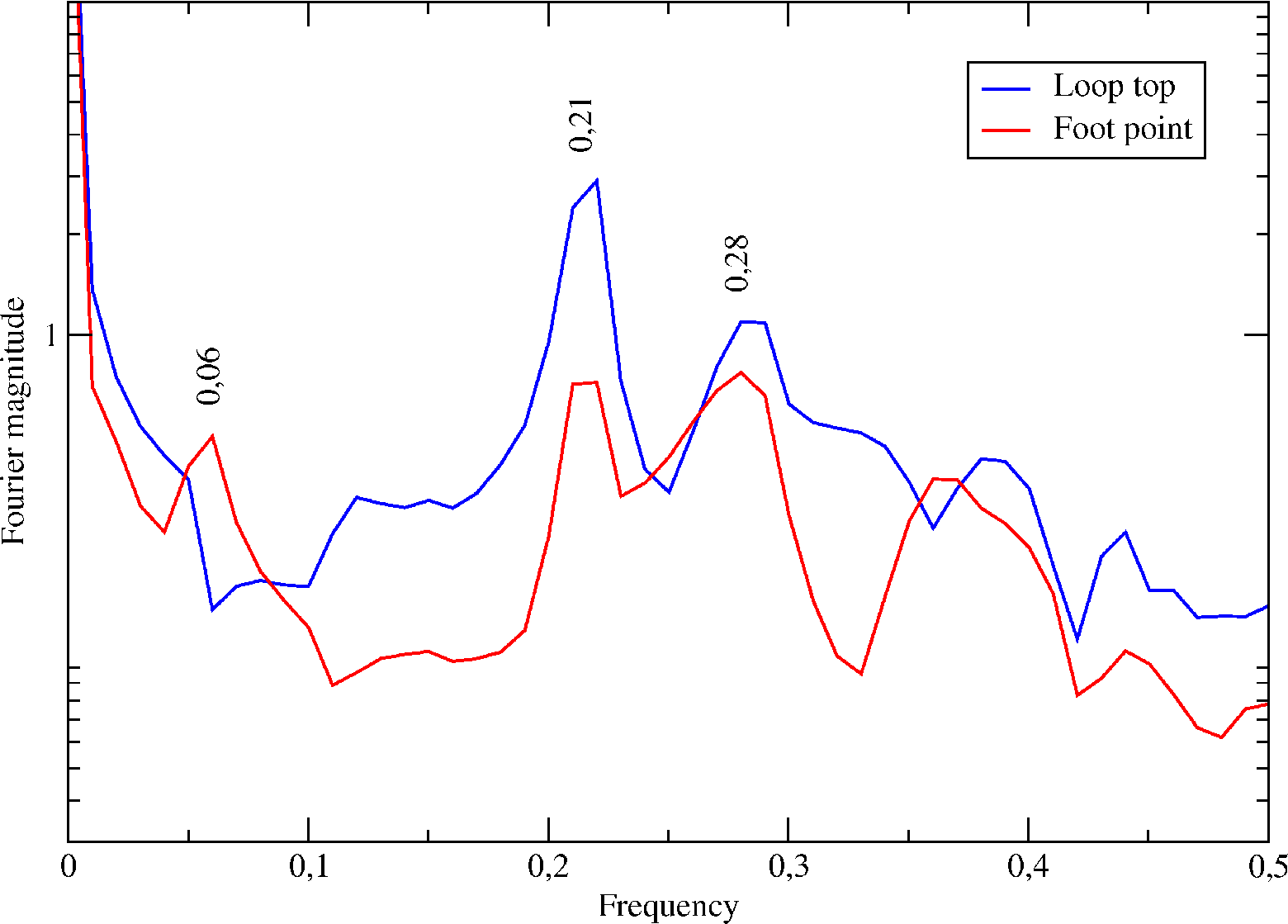}
\caption{Density Fourier Transform  for: Left) Case A, the signal is taken at the grid points $(0.1,0.1)$ and $(0.1,0.5),$ top loop point and footpoint, respectively;  Right) Case M, the footpoint signal taken at the grid point $(0.0,0.5)$ is larger for the second peak. The opposite is true for the loop top signal,  taken at the grid point $(0.0,0.0).$}
  \label{fig:dos}
   \end{center}
\end{figure*}

Figures~\ref{fig:dos}a,b show the  Fourier transform (FT) of the loop density for Case A and M, respectively.

\noindent \textbf{Case A}

\noindent The Case A FT  (Fig.~\ref{fig:dos}a) was calculated at the grid points $(x,y)=(0.1,0.1)$ and $(x,y)=(0.1,0.5)$, at heights of $1$Mm from 
the top loop and $2.5$Mm from the footpoint, respectively. The time is $t=\tau t_{0}$ and the dimensionless frequency is $\nu=1/\tau.$ The 
intense peak of frequency $0.21$ corresponds  to a period of $\sim 6.6$sec which is the fundamental sausage mode described by the observations. 
The frequency $0.34$ correspond to a period of $4.1$sec, associated to the sausage second harmonic.

\begin{figure}
\begin{center}
 \includegraphics[width=7.1cm]{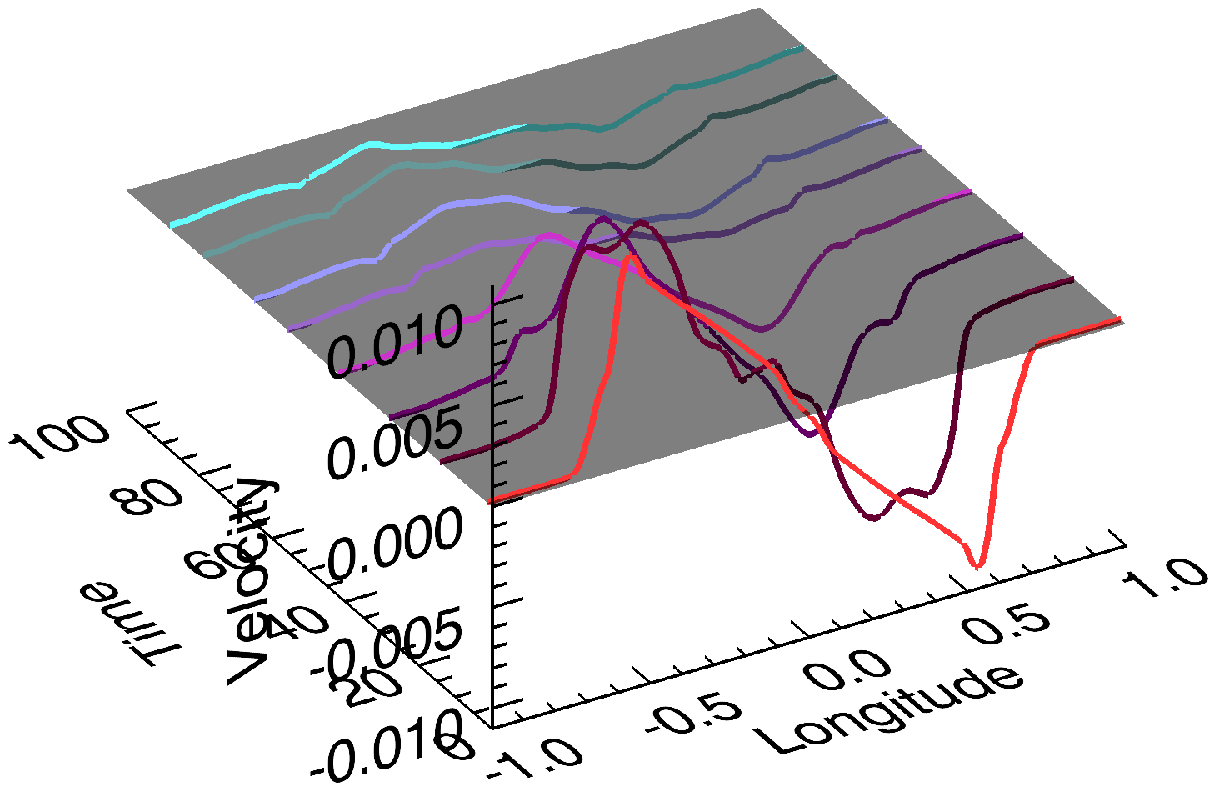}
 \includegraphics[width=7.1cm]{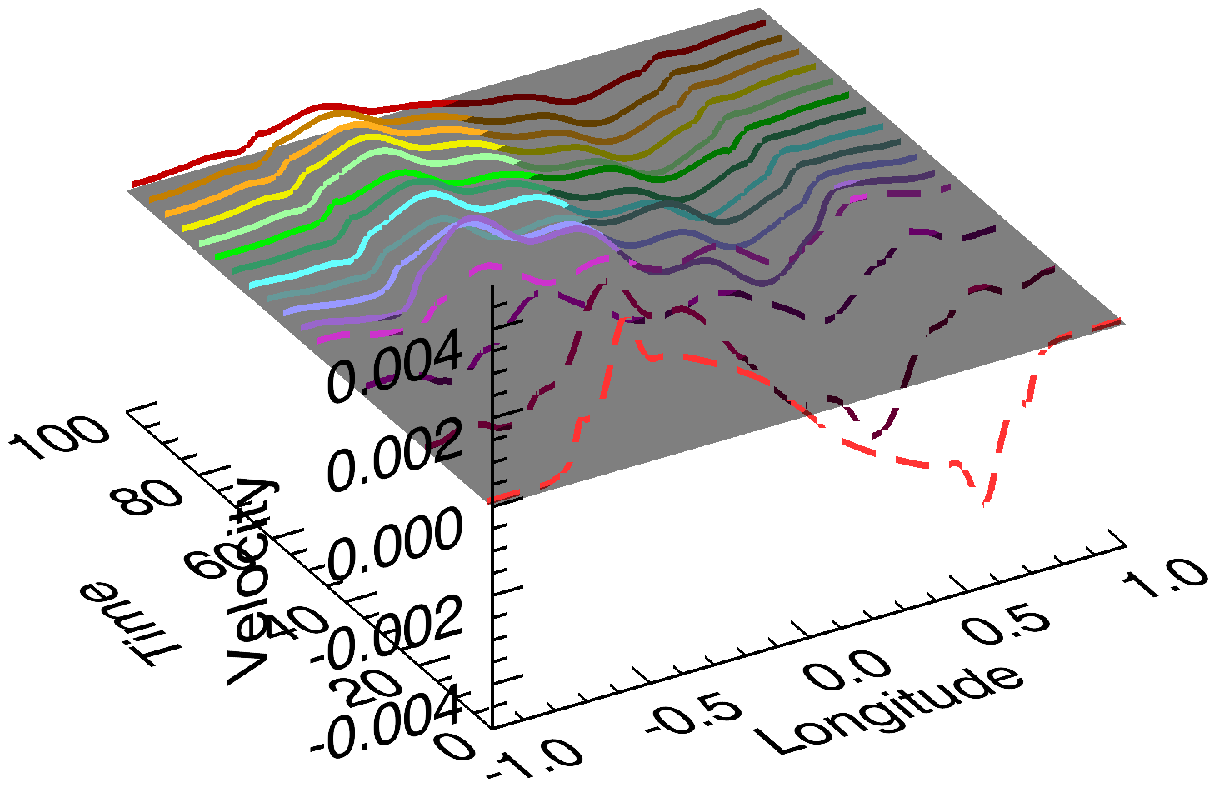}
\caption{$v_{y}$ shape along the loop axis as a function of time. Top) Case A, dimensional quantities obtained multiplying by $l_{0}=10$Mm, $t_{0}=1.39$sec and $v_{0}=7149$Km sec$^{-1}$; Bottom) Case M,  dimensional quantities obtained multiplying by $l_{0}=16.3$Mm, $t_{0}=3.25$sec and $v_{0}=5000$Km sec$^{-1}.$ }
\label{fig:tres}
\end{center}
\end{figure}

\begin{figure}
\begin{center}
\includegraphics[width=7.1cm]{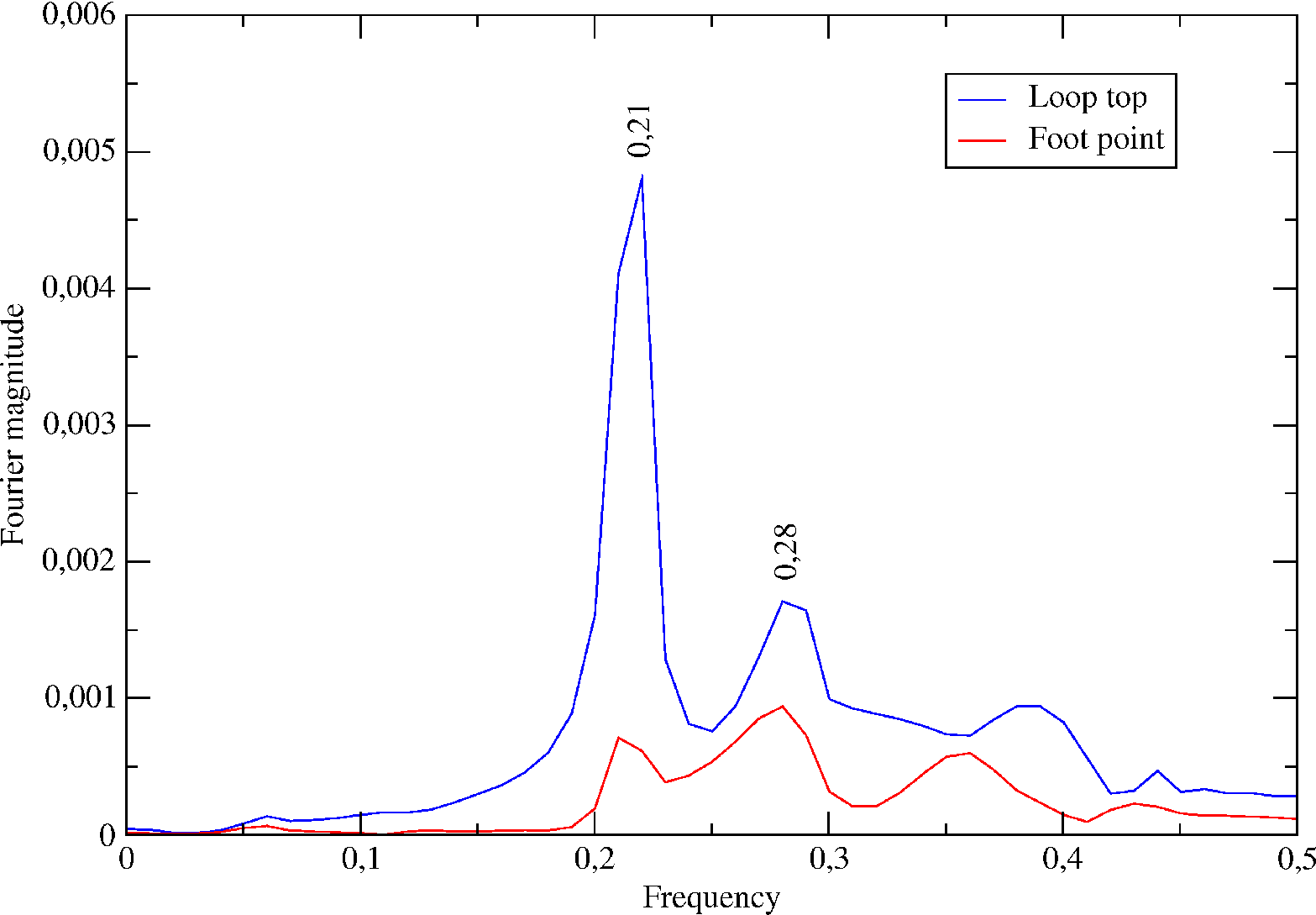}
\includegraphics[width=7.1cm]{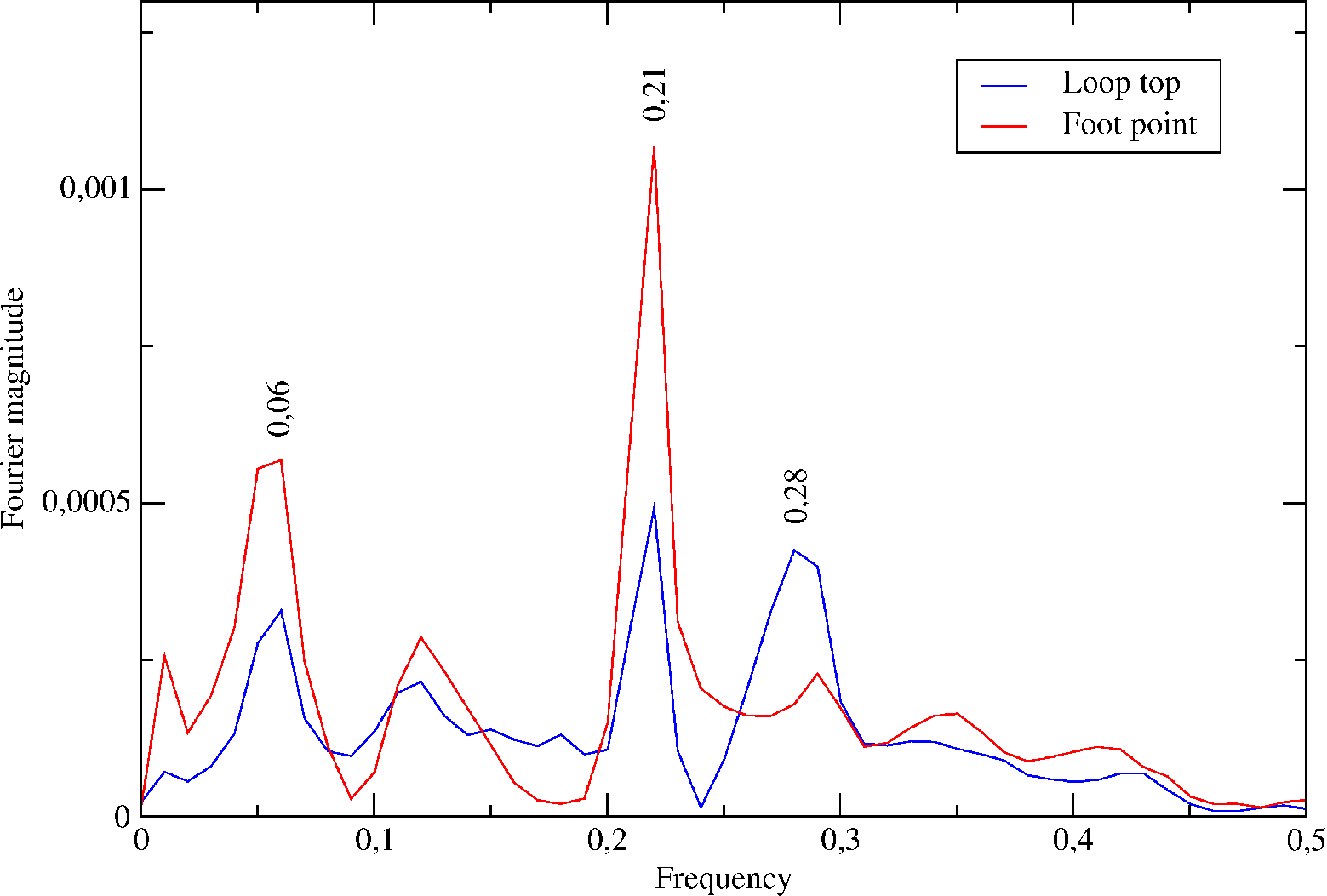}
\caption{Fourier Transform of the Case M signal of: Top) transverse velocity component; Bottom) longitudinal velocity component. The footpoint signal taken at the grid point $(0.0,0.5)$ and the top signal at t he grid point $(0.0,0.1).$ }
\label{fig:cuatro}
\end{center}
\end{figure}

Sausage standing modes in uniform media have wavelengths given by the expression $\lambda=2L/N.$ The  departure from one of the  ratio $P_{1}/2P_{2},$ between the fundamental  ($N=1$) and  second ($N=2$) harmonic, is a measure of the longitudinal stratification imposed by the density contrast between the interior and exterior of the loop (Nakariakov et al. \cite{nak2}; McEwan et al.  \cite{ewa}). In this case 
$P_{1}/2P_{2}=0.8.$ \\
\indent Also, we found that there is a coupling of the sausage mode with slow  components that are  damped approximately in five periods. This is 
shown in Fig.~\ref{fig:tres}a, where the  longitudinal velocity along the loop axis   is displayed. The features are a composition of different modes, all of them with a node at the center of the loop length, the whole pattern resembles a second harmonic one indicating that this mode is predominant, as suggested by Nakariakov et al.  \cite{nak2}. 
The different shapes of Fig.~\ref{fig:tres}a are similar, so we consider  the period as the time distance between two of these curves: $\sim 17$sec. The dimensionless frequency associated to this  period is the 
peak $0.08,$ also shown in Fig.~\ref{fig:dos}a. Away from these maxima the signal gets strongly mixed with other perturbations and it is impossible to see particular features beyond the FT. \\
\noindent \textbf{Case M} \\
\noindent Figure~\ref{fig:dos}b shows the FT spectrum of Case M for the grid points $(x,y)=(0.0,0.0)$ and $(x,y)=(0.0,0.5)$, at the top loop and at a height of   $4.1$Mm from the footpoint, respectively.
Melnikov et al. \cite{mel} and Nakariakov et al. \cite{nak2} found that there are two main spectral components of the observed pulsations, 
with periods $P_{1}=14-17$sec and $P_{2}=8-11$sec. Note that, as in Case A, the  simulation of Case M shows that there are two main  frequencies of  
characteristic periods  $\sim 15.5$sec ($\nu=0.21$) and $\sim 11.6$sec ($\nu=0.28$), respectively. Both of them are registered everywhere in the loop structure. 
However, as these authors stated, there is a difference in the contribution of the peaks depending on the location.  At the legs the frequency  $\nu=0.28$ is as important as  $\nu=0.21.$ On the other hand, at the top, the frequency $\nu=0.21$ has a larger amplitude  than $\nu=0.28.$ 
Nakariakov et al. \cite{nak2} suggested that $\nu=0.28$ corresponds to higher spatial harmonic. 
To discuss this point we performed the FT to the transverse velocity 
signal as an indication of transverse compression; the result is shown in Fig.~\ref{fig:cuatro}a. Note that,  at the footpoints, the 
second peak is the most intense, the opposite occurs at the top. We found that the frequency $0.28$ is the second sausage harmonic mode. In this case $P_{1}/2P_{2}=0.67.$ 
Figure~\ref{fig:tres}b shows the most important slow mode contributions, all of them with a node at the center and a whole pattern resembling a second slow harmonic. There is a differentiated behavior in time. The  first curves (dashed) shown in Fig.~\ref{fig:tres}b are later replaced by others (solid curves) of smaller amplitude and period that last not damped for at least $10$ periods.
This two  main periods  can also be identified in Fig.~\ref{fig:dos}b as the frequencies $\nu=0.06$ ($54$sec) and $\nu=0.21$ ($15.5$sec).   As in Case A, there is a  slow  frequency;  but there is a second one whose  frequency is coincident  with the sausage fundamental one. 
It seems that this  slow contribution is  driven by the fast modes of the same frequency. 
The damping of Case M sausage signal shown in Fig.~\ref{fig:uno}b could be due to the  drain of energy from the sausage mode into the  inner longitudinal perturbations.  In Case A the longitudinal perturbations are not resonantly coupled to the sausage mode and  thus, are not capable to damp the sausage  mode. 
Figure~\ref{fig:cuatro}b displays the Case M FT for the longitudinal velocity component ($v_{y}$). As a difference with Fig.~\ref{fig:cuatro}a, due to the resonant slow mode contribution,  the peak of the  fundamental sausage mode, is higher at the footpoint than at the top. 

 \begin{figure}
\begin{center}
  \includegraphics[width=7.1cm]{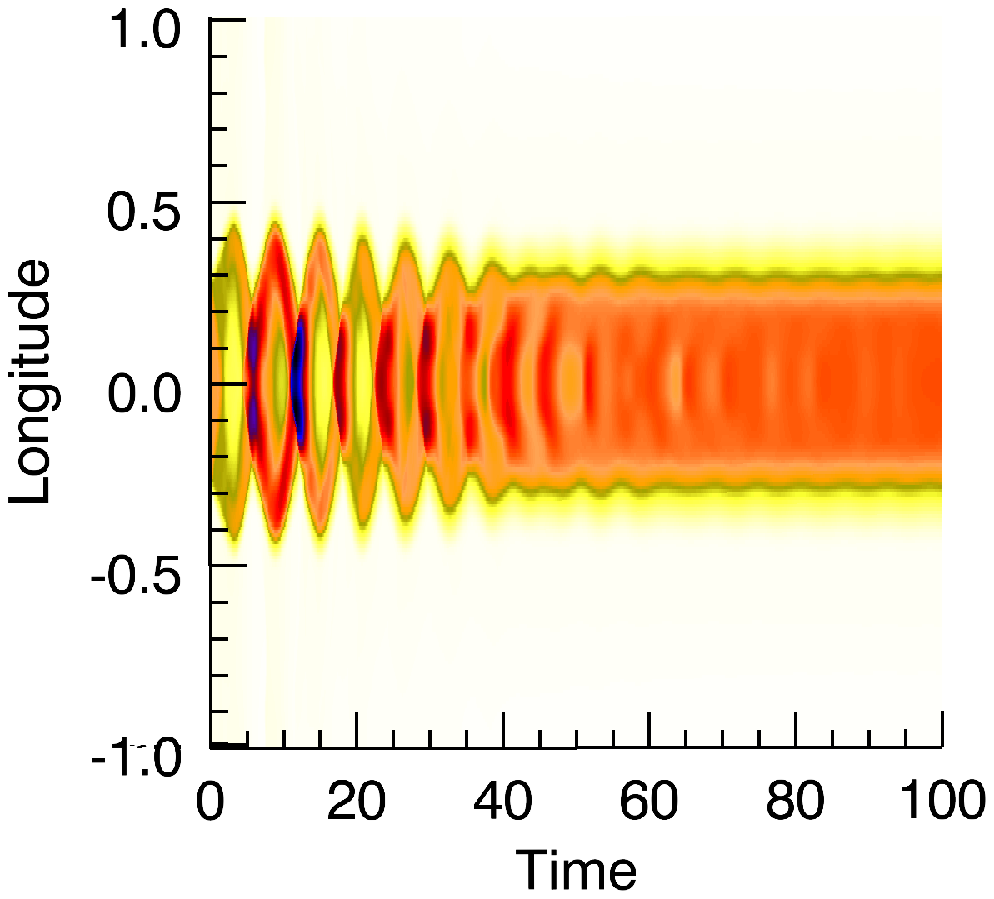}
  \includegraphics[width=7.1cm]{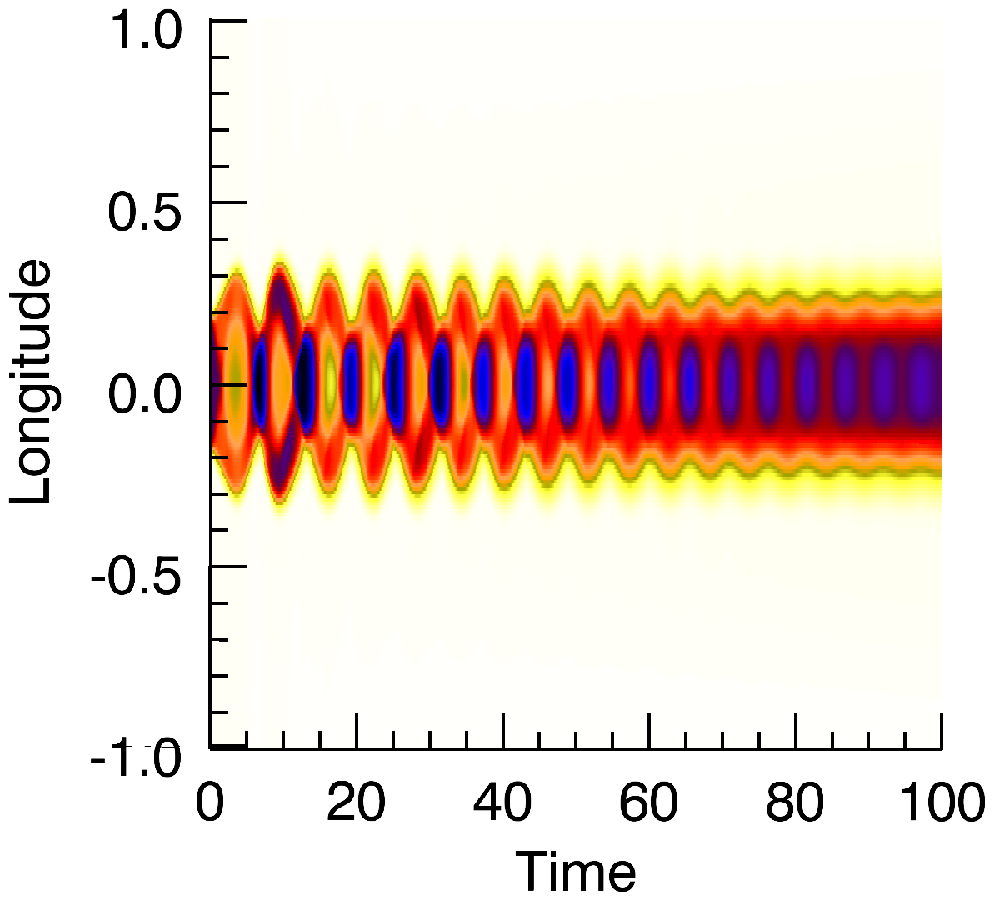}
\caption{Density as a function of time, across the loop width ($x$), and at the center of the loop length, ($y=0$). Top) Case A, and $\Delta p=0.28.$ Dimensional quantities obtained multiplying by $l_{0}=10$Mm and $t_{0}=1.39$sec; Bottom) Case M, and $\Delta p=0.35.$ Dimensional quantities obtained multiplying by $l_{0}=16.3$Mm and $t_{0}=3.25$sec.}
  \label{fig:cinco}
   \end{center}
\end{figure}
\subsection{Plane wave mode decomposition}
To better understand the mode contribution in the observed phenomena, we perform a plane wave  decomposition. This procedure gives limited information, due to the fact the wavelength of the perturbation is, in some cases, of the order of the background quantities, which make difficult to establish the real strength of the mode components. Some quotients between different Fourier  components, supposedly pertaining to the  same mode, have the expected values arising from the calculation of this standard mode decomposition, while others differ. However, this is dependent of the location of the signal analyzed. Thus, we compare the quotients between different field components: those resulting from the eigenvector decomposition into modes, with the quotients coming from the FT. The eigenvectors of the system were computed in previous studies (Pennisi \citealp{penni}, Balsara \citealp{bals}, C\'ecere et al. \citealp{cec}), thus we  use the ratios of the eigenvector components and express them as functions of the background quantities. As in this highly varying dynamics it is difficult to properly define the background quantities, we averaged the values (in time and space) of the different fields near the center of the loop and near the footpoints. 
The comparison between the different mode components $\bar{p}/ \bar{\rho},$ $ \bar{v}_x/ \bar{\rho}, $ $\bar{v}_y/ \bar{\rho}, $ $ \bar{B}_x/ \bar{\rho}, $ $ \bar{B}_y/ \bar{\rho}$ of Table~\ref{tab:table2}, and the quotients obtained by doing a FT of  the fields: $\hat{p}/ \hat{\rho},$ $  \hat{v}_x/ \hat{\rho}, $ $ \hat{v}_y/ \hat{\rho}, $ $ \hat{B}_x/ \hat{\rho}, $ $ \hat{B}_y/ \hat{\rho}$ shown in  Table~\ref{tab:table3}, were performed taking these values at the main characteristic frequencies. This information is extracted, in each case, from the top of the loop and from the footpoints. 

If for each case, we compare Table~\ref{tab:table2} with Table~\ref{tab:table3} for the loop top and for the footpoints, respectively, we find an overall reasonable agreement. However, we see that, for the loop top, the ratios that involve the magnetic and the velocity fields are higher than expected for $\nu=0.34$ in Case A, and for $\nu=0.27$ in Case M. This is, for the second harmonic fast modes. If we compare the tables of the footpoints, the disagreement occurs only in the Melnikov slow mode ($\nu=0.06$). On the other hand, the comparison between quotients of fast modes (both frequencies) shows a good accordance. These results reinforce that the FT analysis performed in last section, gives reliable information. The loop top analysis gives information about the fundamental sausage and the slow modes, and the footpoints analysis about the sausage ones.
\begin{table*}
\begin{center}
\begin{tabular}{|l|c|c|c|c|c|l|c|c|c|c|c|} \hline
\multicolumn{12}{|c|}{AT THE LOOP TOP\rule{0pt}{9pt}} \\ \hline
\multicolumn{6}{|c|}{Asai\rule{0pt}{9pt}} &
\multicolumn{6}{|c|}{Melnikov} \\ \hline
\multicolumn{1}{|c|}{\bf Frequency\rule{0pt}{9pt}} &
\multicolumn{1}{c|}{$\hat{p}/ \hat{\rho}$} &
\multicolumn{1}{c|}{$\hat{v}_x/ \hat{\rho}$}&
\multicolumn{1}{c|}{$\hat{v}_y/ \hat{\rho}$}&
\multicolumn{1}{c|}{$\hat{B}_x/ \hat{\rho}$}&
\multicolumn{1}{c|}{$\hat{B}_y/ \hat{\rho}$}&
\multicolumn{1}{|c|}{\bf Frequency} &
\multicolumn{1}{c|}{$\hat{p}/ \hat{\rho}$} &
\multicolumn{1}{c|}{$\hat{v}_x/ \hat{\rho}$}&
\multicolumn{1}{c|}{$\hat{v}_y/ \hat{\rho}$}&
\multicolumn{1}{c|}{$\hat{B}_x/ \hat{\rho}$}&
\multicolumn{1}{c|}{$\hat{B}_y/ \hat{\rho}$}
\\ \hline
0.08 S\rule{0pt}{9pt} & $0.0147$  & $0.0005$ & $0.003$  & $0.002$ & $0.016$ & 0.06 S\rule{0pt}{9pt} & $0.006$ & $0.0002$ & $0.0006$  & $0.001$ & $0.012$     \\ \hline
0.21 F\rule{0pt}{9pt} & $0.011$   & $0.007$  & $0.001$  & $0.004$ & $0.039$ & 0.21 F\rule{0pt}{9pt} & $0.012$ & $0.006$  & $0.0007$  & $0.004$ & $0.02$      \\ \hline
0.34 F\rule{0pt}{9pt} & $0.016$   & $0.012$  & $0.026$  & $0.024$ & $0.035$ & 0.27 F\rule{0pt}{9pt} & $0.02$  & $0.01$   & $0.0015$  & $0.015$ & $0.015$     \\ \hline
\end{tabular}
\end{center}
\begin{center}
\begin{tabular}{|l|c|c|c|c|c|l|c|c|c|c|c|} \hline
\multicolumn{12}{|c|}{AT THE FOOTPOINTS\rule{0pt}{9pt}} \\ \hline
\multicolumn{6}{|c|}{Asai\rule{0pt}{9pt}} &
\multicolumn{6}{|c|}{Melnikov} \\ \hline
\multicolumn{1}{|c|}{\bf Frequency\rule{0pt}{9pt}} &
\multicolumn{1}{c|}{$\hat{p}/ \hat{\rho}$\rule{0pt}{9pt}} &
\multicolumn{1}{c|}{$\hat{v}_x/ \hat{\rho}$\rule{0pt}{9pt}}&
\multicolumn{1}{c|}{$\hat{v}_y/ \hat{\rho}$\rule{0pt}{9pt}}&
\multicolumn{1}{c|}{$\hat{B}_x/ \hat{\rho}$\rule{0pt}{9pt}}&
\multicolumn{1}{c|}{$\hat{B}_y/ \hat{\rho}$\rule{0pt}{9pt}}&
\multicolumn{1}{|c|}{\bf Frequency} &
\multicolumn{1}{c|}{$\hat{p}/ \hat{\rho}$\rule{0pt}{9pt}} &
\multicolumn{1}{c|}{$\hat{v}_x/ \hat{\rho}$\rule{0pt}{9pt}}&
\multicolumn{1}{c|}{$\hat{v}_y/ \hat{\rho}$\rule{0pt}{9pt}}&
\multicolumn{1}{c|}{$\hat{B}_x/ \hat{\rho}$\rule{0pt}{9pt}}&
\multicolumn{1}{c|}{$\hat{B}_y/ \hat{\rho}$\rule{0pt}{9pt}}
\\ \hline
0.08 S\rule{0pt}{9pt} & $0.009$  & $0.0001$ & $0.009$  & $0.004$ & $0.007$ & 0.06 S\rule{0pt}{9pt} & $0.05$  & $0.0001$ & $0.01$   & $0.02$  & $0.051$     \\ \hline
0.21 F\rule{0pt}{9pt} & $0.01$   & $0.009$  & $0.002$  & $0.02$  & $0.05$  & 0.21 F\rule{0pt}{9pt} & $0.009$ & $0.004$  & $0.0006$ & $0.01$  & $0.023$      \\ \hline
0.34 F\rule{0pt}{9pt} & $0.008$  & $0.01$   & $0.002$  & $0.01$  & $0.04$  & 0.27 F\rule{0pt}{9pt} & $0.01$  & $0.005$  & $0.0009$ & $0.007$ & $0.018$    \\ \hline
\end{tabular}
\end{center}
\caption{\label{tab:table2} Main frequency strengths of the field FT quotients: $\hat X$ represents the FT component of the $X$ variable. Top) Calculated near to the loop top and Bottom) near the footpoints.}
\end{table*}

\begin{table*}
\begin{center}
\begin{tabular}{|l|c|c|c|c|c|c|c|c|c|c|} \hline
\multicolumn{11}{|c|}{AT THE LOOP TOP\rule{0pt}{9pt}} \\ \hline
\multicolumn{1}{|c|}{\rule{0pt}{9pt}} &
\multicolumn{5}{|c|}{Asai} &
\multicolumn{5}{|c|}{Melnikov} \\ \hline
\multicolumn{1}{|c|}{\bf{Modes} \rule{0pt}{9pt}} &
\multicolumn{1}{c|}{$\bar{p}/ \bar{\rho}$} &
\multicolumn{1}{c|}{$\bar{v}_x/ \bar{\rho}$}&
\multicolumn{1}{c|}{$\bar{v}_y/ \bar{\rho}$}&
\multicolumn{1}{c|}{$\bar{B}_x/ \bar{\rho}$}&
\multicolumn{1}{c|}{$\bar{B}_y/ \bar{\rho}$}&
\multicolumn{1}{c|}{$\bar{p}/ \bar{\rho}$} &
\multicolumn{1}{c|}{$\bar{v}_x/ \bar{\rho}$}&
\multicolumn{1}{c|}{$\bar{v}_y/ \bar{\rho}$}&
\multicolumn{1}{c|}{$\bar{B}_x/ \bar{\rho}$}&
\multicolumn{1}{c|}{$\bar{B}_y/ \bar{\rho}$}
\\ \hline
Slow \rule{0pt}{9pt}    & $0.01$ & $0$     & $0.005$ & $0$  & $0$    & $0.008$ & $0$     & $0.001$ & $0$  & $0$    \\ \hline
Fast \rule{0pt}{9pt}    & $0.01$ & $0.009$ & $0$     & $0$  & $0.04$ & $0.008$ & $0.002$ & $0$     & $0$  & $0.01$ \\ \hline
\end{tabular}
\end{center}
\begin{center}
\begin{tabular}{|l|c|c|c|c|c|c|c|c|c|c|} \hline
\multicolumn{11}{|c|}{AT THE FOOTPOINTS\rule{0pt}{9pt}} \\ \hline
\multicolumn{1}{|c|}{\rule{0pt}{9pt}} &
\multicolumn{5}{|c|}{Asai} &
\multicolumn{5}{|c|}{Melnikov} \\ \hline
\multicolumn{1}{|c|}{\bf{Modes} \rule{0pt}{9pt}} &
\multicolumn{1}{c|}{$\bar{p}/ \bar{\rho}$} &
\multicolumn{1}{c|}{$\bar{v}_x/ \bar{\rho}$}&
\multicolumn{1}{c|}{$\bar{v}_y/ \bar{\rho}$}&
\multicolumn{1}{c|}{$\bar{B}_x/ \bar{\rho}$}&
\multicolumn{1}{c|}{$\bar{B}_y/ \bar{\rho}$}&
\multicolumn{1}{c|}{$\bar{p}/ \bar{\rho}$} &
\multicolumn{1}{c|}{$\bar{v}_x/ \bar{\rho}$}&
\multicolumn{1}{c|}{$\bar{v}_y/ \bar{\rho}$}&
\multicolumn{1}{c|}{$\bar{B}_x/ \bar{\rho}$}&
\multicolumn{1}{c|}{$\bar{B}_y/ \bar{\rho}$}
\\ \hline
Slow \rule{0pt}{9pt}    & $0.01$ & $0$     & $0.005$ & $0$  & $0$    & $0.009$ & $0$     & $0.001$ & $0$  & $0$   \\ \hline
Fast \rule{0pt}{9pt}    & $0.01$ & $0.009$ & $0$     & $0$  & $0.04$ & $0.009$ & $0.002$ & $0$     & $0$  & $0.01$    \\ \hline
\end{tabular}
\end{center}
\caption{\label{tab:table3} Quotients between components of the slow and fast eigenvectors: $\bar X$ represents the eigenvector component $X$, calculated with the averaged background values. Top) Calculated near to the loop top and Bottom) near the footpoints.}
\end{table*}
 
\subsection{Non equilibrium pressure across the loop radius}
\begin{figure}
\begin{center}
 \includegraphics[width=7.1cm]{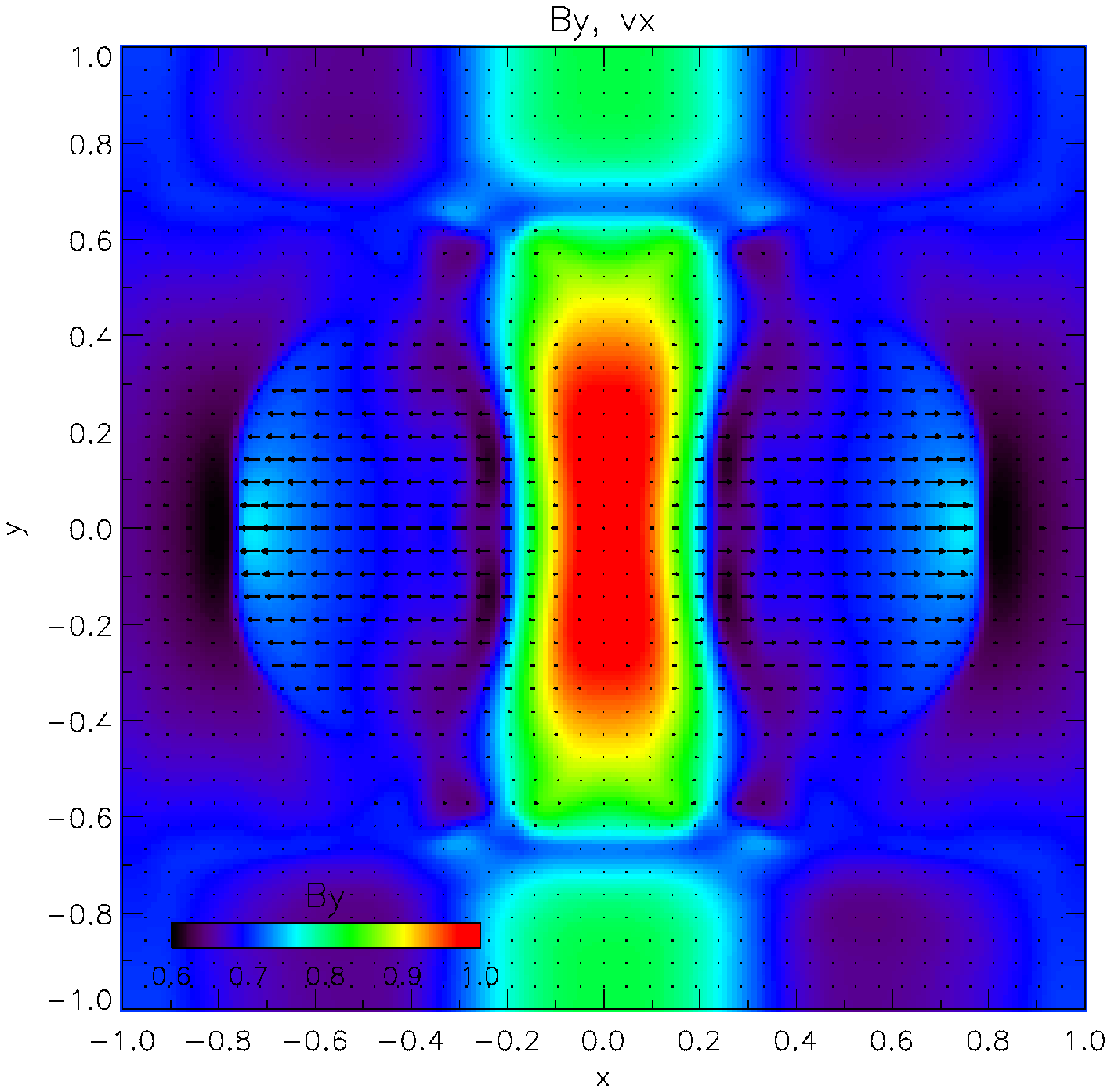}
  \includegraphics[width=7.1cm]{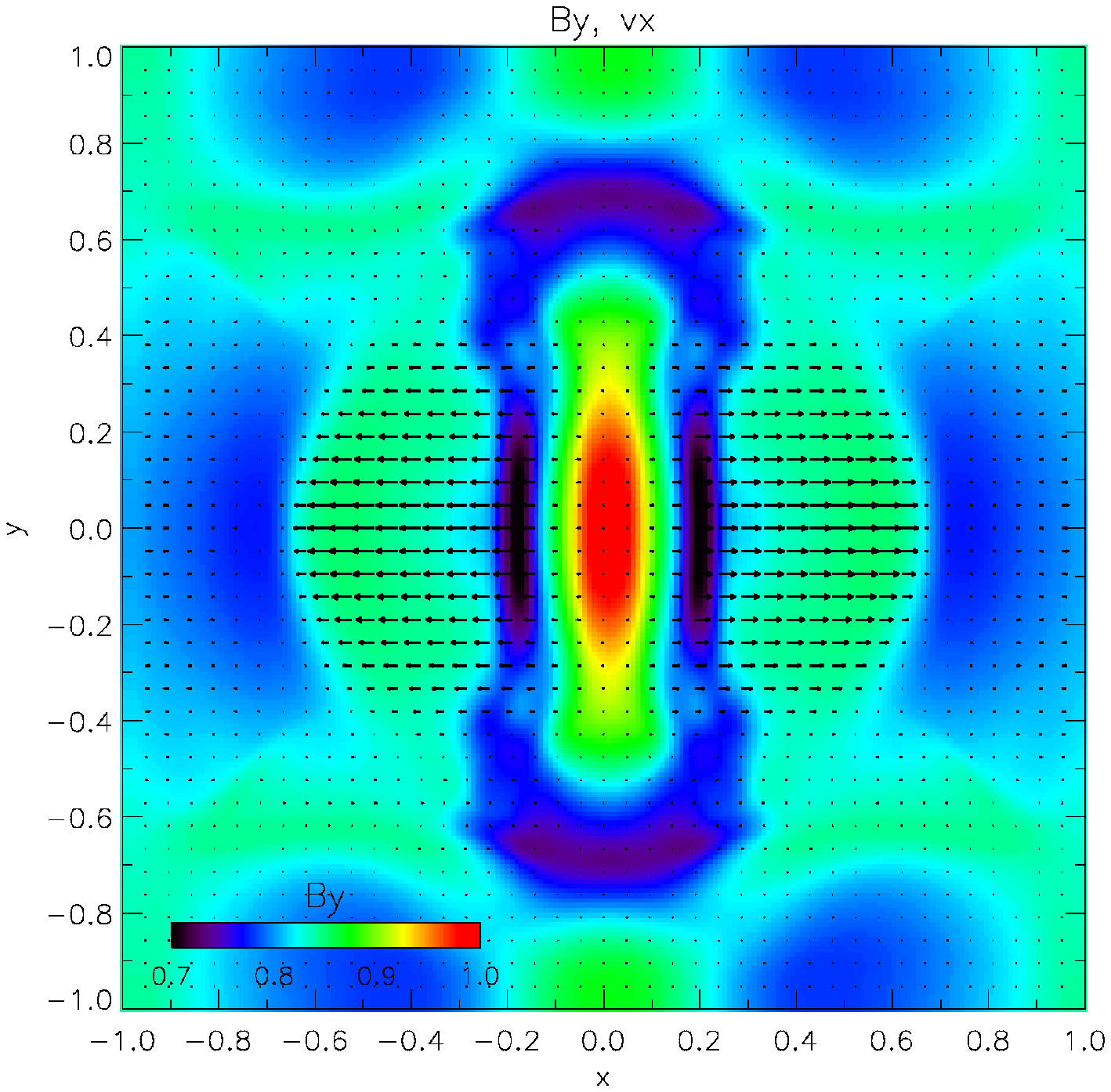}
 \end{center}
\caption{Longitudinal magnetic field component and, superimposed arrows, transverse velocity component. Top) Case A, Bottom) Case M.}
  \label{fig:seis}
\end{figure}

Post--flare loops are part of  a scenario where  impulsive energy bursts are common and become available perturbations that excite oscillations and instabilities in magnetic structures.  Thus, we investigated the response of the loop system to a  perturbation that  induces a discontinuity, i.e., an impulsive release of energy  modelled as a sudden total pressure discontinuity across the loop radius. 
We choose the dimensionless pressure pulses: $\Delta p_{A} = 0.28$ and $\Delta p_{M} = 0.35.$
They are equivalent to a release of energy of $\sim 500$erg cm$^{-3}$ and $\sim 200$erg cm$^{-3}$
which are lower than the typical excess energy content in a flaring loop 
(Kirk et al.\citealp{kirk}). 
 With these Rankine--Hugoniot conditions we obtained   Figs.~\ref{fig:cinco}a,b. The figures show the time evolution of the density oscillations for Case A and Case M, respectively. Note that the oscillations have larger amplitudes than in the correspondent Figs.~\ref{fig:uno}a,b and are rapidly  damped in $\sim 8-12$ periods.
 The fundamental sausage modes have  amplitudes  of  $\sim 83\% R_{l},$  in Case A and $\sim 100\% R_{l},$  in Case M.  
 The damping occurs because the discontinuity imposes a transformation of the trapped modes into leaky modes.   Figures~\ref{fig:seis}a,b show a slice of the magnetic field longitudinal component where the superimposed  arrows represent the transversal velocity component for Case A and Case M, respectively,  at the time $t=26$sec  and $t=44$sec.  As the magnetic field and the velocity are  frozen in quantities, the arrows are tracers of the energy  drained away from the loop.  At the center of the external region a compressional  Alfv\'en wave is excited  and the perturbed magnetic energy of the loop is damped. The excitation at the footpoints is negligible due to the change in the $\beta$ parameter and the correspondent line--tied to the chromospheric bases. 

\section{Conclusions}
We integrated the ideal MHD equations to simulate fundamental trapped sausage modes observationally described  (Asai et al. \citealp{asa} (Case A) and Melnikov et al. \citealp{mel} (Case M)) incorporating a dense chromospheric region.  We could reproduce the observational parameters described in the literature, i.e, periods, densities, lengths and magnetic fields.
As in Melnikov et al. \cite{mel}, we found that, for both cases, there are two  peak fast frequencies with different contributions depending on the location of the signal. The modes  were interpreted as the fundamental (more intense at the top of the loop) and the  second sausage harmonic (more intense at the footpoints) (see Fig.~\ref{fig:dos}a for Case A and  Fig.~\ref{fig:dos}b and Fig.~\ref{fig:cuatro}a for Case M).  
We analyzed the coupling of the fundamental sausage  modes with  longitudinal components. In both cases we obtained  slow frequency components, of $P\sim 17$sec and $P\sim 54$sec, respectively. Moreover, we found that Case M has another   slow contribution, of the same frequency as the fundamental  sausage mode,  apparently driven by this frequency.
The slow  component of Case A is rapidly damped ($\sim 5$ periods); whereas the -more broad slow spectra- of Case M  lasts not damped for various periods. We suggest that this is due to the internal transferring of the fast mode energy into the slow energy which is accomplished more efficiently in the resonant case. Thus, we call this mechanism  an internal damping one. \\
\indent We also  showed that certain initial conditions  associated with  typical flaring loop energy releases (impulsive depositions of energy) that could be achieved by considering jump conditions across the radio, results in the drain of energy into the exterior in the form of a leaky compressible Alfv\'en mode, allowing the final damping of the initially trapped fundamental sausage mode.   This   external damping mechanism which is efficient to damp  modes in an initially trapped configuration is in accordance with typical observational damping times of this modes ($\leqslant 10$periods).

\section*{Acknowledgments}

We would like to thank Carlos Palenzuela for some helpful suggestions. This work was supported in part by SECyT-UNC, CONICET and ANPCyT.

\end{document}